\input harvmac
\newcount\figno
\figno=0
\def\fig#1#2#3{
\par\begingroup\parindent=0pt\leftskip=1cm\rightskip=1cm
\parindent=0pt
\baselineskip=11pt
\global\advance\figno by 1
\midinsert
\epsfxsize=#3
\centerline{\epsfbox{#2}}
\vskip 12pt
{\bf Fig. \the\figno:} #1\par
\endinsert\endgroup\par
}
\def\figlabel#1{\xdef#1{\the\figno}}
\def\encadremath#1{\vbox{\hrule\hbox{\vrule\kern8pt\vbox{\kern8pt
\hbox{$\displaystyle #1$}\kern8pt}
\kern8pt\vrule}\hrule}}

\overfullrule=0pt

\Title{{\it TIFR-TH/97-21}}
{\vbox{\centerline{Emission from parallel $p$-brane Black Holes}}}
\smallskip
\centerline{Sumit R. Das\foot{E-mail: das@theory.tifr.res.in}}
\smallskip
\centerline{\it Tata Institute of Fundamental Research}
\centerline{\it Homi Bhabha Road, Bombay 400 005, INDIA}
\smallskip
\bigskip

\medskip

\noindent

The entropy of a near-extremal black hole made of parallel D-branes
has been shown to agree, upto a numerical factor, with that of the gas
of massless open string states on the brane worldvolume when the
string coupling is chosen suitably. We investigate the process of
emission or absorption of massless S-wave neutral scalars by these
black holes. We show that with rather mild assumptions about the
nature of the interactions between the scalar and open string states,
the D-brane cross-section generally fails to reproduce the universal
low energy black hole cross-section except for 1-branes and 3-branes.

\Date{May, 1997}
\lref\susskind{L. Susskind, hep-th/9309145;
J. Russo and L. Susskind, Nucl. Phys. B 437 (1995) 611}
\lref\duff{ M. Duff and
J. Rahmfeld, Phys. Lett. B345 (1995) 441, hep-th/9406105}
\lref\sen{ A. Sen,
Nucl. Phys. B440 (1995) 421 , hep-th/9411187 and Mod. Phys. Lett.
A10 (1995) 2081.}
\lref\stromvafa{A. Strominger and C. Vafa, {\it Phys.
Lett.} B379 (1996) 99,
hep-th/9601029.}
\lref\callanmal{C. Callan and J. Maldacena, Nucl. Phys. B 475 (1996) 645,
hep-th/9602043.},
\lref\horostrom{G. Horowitz and A. Strominger, Phys. Rev. Lett.
77 (1996) 2368, hep-th/9602051.}
\lref\dmw{A. Dhar, G. Mandal and S.R. Wadia, 
Phys. Lett. B388 (1996) 51, hep-th/9605234.}
\lref\dmatb{S.R. Das and S.D. Mathur, Nucl. Phys. B478 (1996) 561,
hep-th/9606185}
\lref\dmatc{S.R. Das and S.D. Mathur,
 Nucl. Phys. B482 (1996) 153, hep-th/9607149.}
\lref\malstrom{J. Maldacena and A. Strominger, Phys. Rev. D55 (1997) 861,
hep-th/9609026.}
\lref\gubkleb{S. Gubser and I. Klebanov, Phys. Rev. Lett. 77
(1996) 4491, hep-th/9609076.}
\lref\callanfixed{C.Callan, S. Gubser, I. Klebanov and
A. Tseytlin, hep-th/9610172.}
\lref\horeview{For a recent review and other references see
G. Horowitz, gr-qc/9704072.}

\lref\malst{J. Maldacena and
A. Strominger, hep-th/9702015} 
\lref\mathur{S.D. Mathur, hep-th/9704156}
\lref\gubser{S. Gubser, hep-th/9704195}

\newsec{Introduction}
Recently the success of the idea that string states behave as black
holes at strong coupling has revolutionized our understanding of black
hole thermodynamics \refs{\susskind , \duff , \sen , 
\stromvafa , \callanmal ,
\horostrom , \dmw ,\dmatb , \dmatc , \malstrom , \gubkleb ,
\callanfixed , \horeview}.

When the product of the string coupling and the charge is large enough
D-brane configurations are expected to behave as black holes. For
D-brane bound states which correspond to five dimensional extremal
black holes with large horizons, the black hole entropy is indeed
reproduced exactly \stromvafa, a result which is understood from the
non-renormalization of BPS states. The entropy continues to agree for
slight departures from extremality \refs{\callanmal ,\horostrom} and
the very low energy decay rate was found to be proportional to the
horizon area \callanmal\ which is consistent with the semiclassical
Hawking radiation from these holes \dmw. Remarkably the two emission
(or absorption) rates were shown to be {\it exactly} equal
\refs{\dmatb , \dmatc} and even the greybody factors agree exactly
\malstrom\ in the dilute gas regime \foot{Such detailed agreements are
not present beyond the dilute gas regime \ref\klebamat{I. Klebanov and
S.D. Mathur, hep-th/9701187} \ref\traschen{F. Dowker, D. Kastor and
J. Traschen, hep-th/9702109.}}. 
Recent results for such black holes seem to suggest,
however, that there could be enough low energy non-renormalization
properties which justify extrapolation of weak coupling results to the
strong coupling regime \ref\Maldacenab{J. Maldacena, hep-th/9611125},
\ref\das{S.R. Das, hep-th/9703146.}. Most of these results have been
extended to four dimensional holes and for charged scalar emission
\foot{There seem to be some disagreements for emission of charged scalars
when individual energies are large \traschen.}
\ref\four{For references see
\horeview.} and agreements upto numerical factors have been obtained
for emission at higher angular momentum \refs{\malst , \mathur ,
\gubser}.

For parallel D-branes, the extremal state corresponds to black holes
of zero size and one does not expect to reproduce the black hole
entropy exactly upto numerical coefficients, though one expects the
D-brane entropy to be proportional to the area of the stretched horizon
like NS-NS holes or corresponding holes in heterotic string theory
\sen. With a slight departure from extremality the horizon area
becomes non-zero and along the lines of the proposal of \susskind\ one
should be able to understand the entropy after a proper treatment of
mass renormalization of these states. Indeed, as a part of their
general analysis of non-extremal holes in string theory, Horowitz and
Polchinski \ref\horopol{G. Horowitz and J. Polchinski,
hep-th/9612146.}  have shown that if the string coupling $g$ is chosen
such that the curvature (in string metric) at the horizon is of the
order of the string scale, the entropy of a slightly non-extremal
black hole made of $N$ parallel D-branes is exactly reproduced by that
of the gas of the $N^2$ massless modes of the open strings \foot{Note
that this is in a different regime than the work of
\ref\dmata{S.R. Das and S.D.  Mathur, Phys. Lett. B375 (1996) 103.}
where the nonextremal entropy was shown to arise from the excitations
of a single long string whose degrees of freedom are essentially the
$N$ diagonal components of the adjoint fields.}.  When $gN$ is much
smaller than this matching value the horizon curvature is strong, the
semiclassical black hole description is bad while the description in
terms of perturbative D-branes is good. When $gN$ is large the horizon
curvature is small, rendering a black hole description reliable, while
the D-branes become strongly coupled and difficult to describe.  The
above result shows that at the matching point there is no sharp
discontinuity of the mass.

This ``correspondence principle'' also predicts the correct entropy in
the regime far from extremality where for small $gN$ the entropy is
described by that of a single long string \horopol.  In this paper we
will, however, restrict our attention to the near-extremal limit.

It is natural to ask whether the D-brane configuration also absorbs
and emits like a black hole at this matching point.  As mentioned
above it is not completely clear why absorption should work even for
D-brane bound states with a well-understood extremal
entropy. Nevertheless it is important to get as much ``phenomenolgy''
as possible. This is particularly relevant in view of the suspicion
that non-renormalization theorems are at work - these
non-renormalization properties should hold for all slightly
non-extremal parallel D-branes
\das. In fact the S-wave and P-wave absorption by {\it extremal} 
3-branes have been shown to agree {\it exactly} with the classical
answers, \ref\igora{I. Klebanov, hep-th/9702076; S. Gubser, I. Klebanov
and A. Tseytlin, hep-th/9703040.}
presumably because of the non-singular nature of these black
holes.

In the following we calculate the emission cross-section of massless
neutral scalars by such parallel D-branes with zero net momentum by
making rather mild assumptions about the nature of interactions of the
bulk scalars with the open string states. In particular, we allow
processes in which an arbitrary number of open string states can go
into such a scalar via some local interaction at any order of the
open string perturbation theory. We find that at the
lowest energies the D-brane cross-section does not in general
reproduce the universal black hole answer which is the horizon area
\ref\dgm{S.R. Das, G. Gibbons and S.D. Mathur, Phys. Rev. Lett. 78
(1997) 417.}, except for 1-branes and 3-branes. For 1-D branes the
leading process is that of two open strings going into a closed string
while for the 3 D-brane the leading process involves four open strings
going into a closed string.  Our results should also apply to branes
with some net momentum.

\newsec{Non-extremal entropy}

A $p$-brane with RR charge $N$ has a classical solution with the ten
dimensional string metric \ref\horstrom{G. Horowitz and A. Strominger,
Nucl. Phys. B360 (1991) 197.}
\eqn\one{ds^2 = f^{-1/2}[-(1 - {r_0^n \over r^n})dt^2 + dy^i dy_i]
+ f^{1/2}[(1-{r_0^n \over r^n})dr^2 + r^2 d\Omega_{n+1}]}
where $n \equiv 7 - p$, $y^i$ are the coordinates along the brane which is
assumed to be compactified on a torus of volume $V_p$ and $r$ denotes the
radial coordinate in the $9-p$ dimensional noncompact space. 
The harmonic function $f(r)$ is given by
\eqn\two{f(r) = 1 + {r_0^n \sinh^2 \alpha \over r^n}}
Other properties of the classical solution may be found in \horstrom\ and
\horopol. The charge $N$ is given by
\eqn\three{ N \sim {r_0^n \over g} \sinh ~2\alpha}
In \three\ we have used string units and will continue to do so
in the rest of the paper. $g$ is the string coupling.
The horizon is at $r = r_0$ and the classical Beckenstein-Hawking entropy
is given by
\eqn\four{S_{BH} \sim {r_0^{n+1} V_p \over g^2} \cosh~\alpha}
The extremal limit is $r_0 \rightarrow 0$ and $\alpha \rightarrow \infty$
with $N$ held fixed. In this limit $S_{BH}$ is zero.
The total energy is given by
\eqn\fourt{E \sim {r_0^n V_p \over g^2}[{n+2 \over n} + \cosh ~ 2\alpha]}

The curvature at the horizon becomes of the order of the string scale
when \horopol
\eqn\five{r_0 \sim (\cosh ~ \alpha)^{-1/2}}
in string units, and at this point the effective open string coupling
$gN$ becomes
\eqn\six{gN \sim (\sinh ~2\alpha )(\cosh ~ \alpha)^{-n/2}}

For the near-extremal situation (large $\alpha$), 
$r_0 \sim e^{-\alpha/2}$ and
\eqn\mone{gN \sim r_0^{n-4}~~~~~~~~~~~~~S_{BH} \sim {r_0^{n-1} V_p \over
g^2}}
so that the area of the horizon $A_H$ is
\eqn\mtwo{ A_H \sim r_0^{n-1}}
at the matching point.

It was shown in \horopol\ that at this value of the string coupling
the entropy $S_{BH}$ for a near extremal hole with large $\sigma$ 
agrees (upto a numerical factor) with the entropy
of the gas of $N^2$ massless open string modes
\eqn\seven{ S_g \sim N^2 V_p T^p}
where $T$ is the temperature, when the energy of this gas 
\eqn\sevent{\Delta E \sim N^2 V_p T^{p+1}}
is set equal to the {\it excess} 
energy of the corresponding black hole above
extremality.  Equation \seven\ shows that at this matching point the
temperature at this matching point is
\eqn\eight{ T \sim r_0}
As explained in \horopol\ it is immaterial whether we use the asymptotic
or horizon values of $T$ and $V_p$ in \seven\ since the redshift
factor cancels in this equation. 

For near-extremal 3-branes the entropy of the gas agrees with the
black hole entropy (upto a constant) for {\it any value of the string
coupling constant} \horopol, as first noticed in 
\ref\igorpeet{S. Gubser, I. Klebanov and A. Peet, Phys. Rev. 
D54 (1996) 3915, hep-th/9602135.; I. Klebanov and A. Tseytlin,
Nucl. Phys. B 475 (1996) 165, hep-th/9604089.}.  This may be easily
seen by equating \sevent\ with the excess energy obtained from \fourt\
to determine $T$ in terms of $r_0$ and $\alpha$ (after using \six\ to
relate $gN$ to $r_0$ and $\alpha$) and substituting the result in
\seven. The answer agrees with $S_{BH}$ precisely when $p = 3$.

\newsec{Absorption cross-section}

From the point of view of the theory of massless open string modes on
the $p$-brane worldvolume, this slightly nonextremal state decays by
some number of open string modes colliding to go into a massless
closed string state which we take to be a scalar (from the $10-p$
dimensional point of view).  Consider $n_o$ such open string modes with
momenta $p_i~, i = 1,\cdots n_o$ going into a closed string mode with
momenta $(q,k)$ where $q$ denotes the momentum components along the
brane worldvolume and $k$ denotes the components in the noncompact $d=
9-p$ dimensonal space with volume $V_d$.

We will consider a completely general interaction with the form factor
$f(p_i, \omega)$.  We allow some of the initial open string modes to
be fermionic: let there be $n_B$ bosonic and $n_F$ fermionic open
string modes, $n_o = n_B + n_F$. 
$n_F$ is even since the final state is bosonic. If the process 
occurs at the tree level the matrix element is given by
\eqn\nine{M \sim g^{n_o/2} \delta^{p+1}(\sum p_i - q)[\prod_{i=1}^{n_B}
{1 \over (|p_i| V_p)^{{1 \over 2}}}] [\prod_{i=1}^{n_F}{1 \over V_p^{{1
\over 2}}}] ({1 \over \omega V_p V_d})^{{1\over 2}} f(p_i,k)}
To obtain the decay rate we have to average over all initial states
which are taken from a thermal bath at temperature $T$ 
\foot{Once again the redshift factor cancels since only ratio of the
energy to the temperature appear in the distribution functions} 
 and sum over all the final states
\foot{ The net
momentum is zero so that there is no chemical potential for the
momentum}. Thus there is a factor of
the Bose-Einstein distribution function $\rho_B (|p_i|, T)$ for each
of the bosonic open string modes and a fermi distribution function
$\rho_F(|p_i|,T)$ for each fermionic mode.  This yields a decay rate
\eqn\ten{\Gamma \sim N^n \int [\prod_{i=1}^{n_B} V_p d^p p_i~
\rho_B(|p_i|,T)] [\prod_{i=1}^{n_F} V_p d^p p_i~\rho_F(|p_i|,T)]~
|M|^2 ~[V_d d^d k]}
The factor of $N^n$ comes from summing over the ``colors'' of the
worldvolume fields which are in the adjoint representation of $U(N)$.

We will consider wavelengths larger than all other length scales in the
problem. In particular the energy of the outgoing mode $\omega$ is
taken to be much smaller than the temperature $T$. In this regime we
can approximate the distribution functions by
\eqn\eleven{\rho_B (p_i,T) \sim {T \over |p_i|}~~~~\rho_F (p_i,T) \sim 
{1 \over 2}}
The dependence of $\Gamma$ on $\omega$ may be then read off by simply
counting the powers of momenta. The form factor will contribute some
power of $\omega$ which we take to be
\eqn\twelve{ f \sim \omega^{M}}
$M$ is then basically the number of derivatives in the position space
interaction term. A local interaction means that $M$ is nonnegative and
integer. Thus the low energy decay rate becomes
\eqn\twelve{\Gamma \sim (gN)^{n_o} \omega^{\beta - 1} T^{n_B} [d^d k]}
where
\eqn\twelvec{ \beta = (n_B + n_F)(p-2) + 2n_F + 2M - (p+1)}
Since the outgoing particle is a boson we have to divide by one factor of
the Bose distribution function and the phase space volume to obtain the
absorption cross-section
\eqn\twelvea{\sigma \sim (gN)^{n_o} \omega^{\beta} T^{(n_B-1)}}

In the above we assumed that the process occurs at the tree level of
the open string theory. It is straightforward to extend the result to
the situation where the process appears at the loop level. If the
number of holes in the worldsheet diagram is $h+1$ one has an
additional overall factor of $(gN)^{2h}$ so that \twelvea\ becomes
\eqn\twelveb{\sigma_h \sim (gN)^{n_o+2h} \omega^{\beta} T^{(n_B-1)}}

We now compare this D-brane answer for the absorption cross-section
to the classical answer which at low energies is given by \dgm\
\eqn\fourteen{\sigma_{cl} = A_H}
regardless of the type of black hole we are considering. By the
correspondence principle of \horopol\ we expect that these two
cross-sections can possibly match only at the value of the coupling
where the entropies match. At this point \mtwo\ and \eight\ imply 
\eqn\fifteena{\sigma \sim \omega^\beta A_H^\gamma}
where $\beta$ is given by \twelvec\ and 
\eqn\sixteen{\gamma =  {(n_B -1) +(n_B + n_F+ 2h)(3-p) \over 6-p}}
The case $p = 6$ is special and will be dealt with later.

The D-brane cross-section matches the classical answer when
\eqn\seventeen{\beta = 0~~~~~~~~~\gamma = 1}
which has the following solutions for $n_B$ and $n_F$
\eqn\eighteen{\eqalign{& n_B = {6M -3 +p(5-2M) 
+ 2hp(p-3) \over 6-p} \cr
& n_F = {18-8M + 2p(M-3) - 2h(p-3)(p-2) \over 6-p}}}
We have to find solutions for these equations for given $p$ with
positive integer values for $n_B, n_F, h$ and $M$. Furthermore since
the minimal process has to involve at least two open strings
$n_o = n_B + n_F \geq 2$ and since the final state is a boson $n_F$
must be even.

Remarkably these conditions are rather restrictive and by a case
by case study it is easily seen that there are precisely two solutions
\eqn\nineteen{\eqalign{& p = 1~~~~~~~n_B = 2~~~n_F = 0~~~M = 2~~~~h=0\cr
& p = 3~~~~~~~n_B = 4~~~n_F = 0~~~M = 0~~~~h=0}}

For the 6-brane the D-brane absorption cross-section becomes
\eqn\twenty{\sigma_{p=6} \sim r_0^{-(3n_F + 2n_B + 6h +1)}\omega^{4n_B
+ 6n_F + 2M - 7}}
In this case one has $n = 1$ and the horizon area or the classical
entropy does not depend on $r_0$ ( or eequivalently $\alpha$).
Thus for \twenty\ to agree with the classical answer the power
of $r_0$ has to be zero. This, however, cannot happen since $n_B$
is strictly positive and $n_F$ is non-negative.

The two cases where the D-brane answer does agree with the classical
answer upto a numerical factor correspond to familiar terms in the
tree level worldbrane action of a $U(N)$ gauge theory. For the
1-brane, the relevant interaction is the familiar term of the
form $\phi \partial X \partial X$ where $\phi$ is the dilaton and
$X$ stands for a transverse coordinate. For the 3-brane the interaction
term involves the commutator term in the gauge field on the brane, e.g.
\eqn\twentyone{{\rm Tr}~\phi~[A_\alpha, A_\beta][A^\alpha, A^\beta]}
or coupling of the ten dimensional graviton longitudinal to the brane
to the energy momentum tensor.

\newsec{Discussion}

The spirit of the calculation of absorption cross-section for D-brane
bound states with large horizons in the extremal limit or for extremal
3-branes is rather different from the considerations of this paper. In
the former situations one compares the tree level D-brane result with
the first term in the expansion of the classical cross-section in
powers of string cooupling \igora, \das. Such an expansion is possible
because of low energies. Then the non-trivial point to check is that
there are no open string loop corrections to the tree level D-brane
answer at the same order of the energy \das. In this paper we tried to
compare D-brane and black holes at a specified value of the coupling,
at which the horizon curvature becomes of the order of the string
scale. In this sense it is not surprising that the cross-sections do
not agree, even though the entropy does.

It is nevertheless interesting to note that the two cases which lead
to agreement (apart from numerical factors) are also
essentially cases which lead to exact agreement of cross-sections
obtained so far. The low energy degrees of freedom of the five
dimensional black hole are in fact those of a long string. Absorption
of S-wave scalars by {\it extremal} 3-branes have been shown to agree
exactly with the classical answer \igora. A puzzling point is that for
absorption by extremal 3-branes (which have of course zero horizon
area) the leading contribution in the D-brane cross-section comes from
the interaction of two open strings with a closed string and leads to
the correct answer proportional to $\omega^3 R^8$ where $R$ is the
gravitational radius. For a non-extremal brane it is the four point
coupling which gives the leading contribution proportional to the area,
while the term involving two open strings behaves as $\omega^2 T$ for
$\omega << T$.

In Type IIB theory 3-branes are special and provide a good laboratory
to understand the D-brane- black hole connection, since the extremal
geometry is nonsingular and the dilaton is a constant. In fact for
this case the entropy of near-extremal D-branes agree with the black
hole entropy upto numerical factors {\it regardless of the value of
the coupling} \igorpeet.
On the other hand we found that the absorption
agrees only at the matching point required by the correspondence
principle which happens when $gN \sim 1$ and independent of $r_0$. 
The significance of this is not clear.

After completion of this work we received a preprint \ref\emp{R. Emparan,
hep-th/9704204} where it is shown that the grey body factors for
NS-NS holes are not correctly predicted from the corresponding string
calculation, though the leading area dependence is recovered. The fact
that the 1D-brane cross-section also reproduced the leading result is
consistent with this since the effective action for emission of quanta
from excited fundamental strings is rather similar to the D-string
effective action \dmatc.

\newsec{Acknowledgements} I would like to thank Samir Mathur for
enlightening discussions and for comments on the manuscript, Saumen
Dutta for a discussion and the Physics Departments of Stanford
University, Rutgers University and Washington University for
hospitality during the course of this work.

\listrefs
\end